\documentclass[aps,prb,showpacs,twocolumn,superscriptaddress,floatfix]{revtex4-1}
\usepackage{hyperref}
\usepackage{graphicx}
\usepackage{amsmath,amssymb,latexsym,amsfonts,subfigure,bbm}
\usepackage{dsfont,relsize,color}

\newcommand{\be}{\begin{equation}}
\newcommand{\ee}{\end{equation}}
\newcommand{\beqn}{\begin{eqnarray}}
\newcommand{\eeqn}{\end{eqnarray}}

\newcommand{\bml}{\begin{multline}}
\newcommand{\eml}{\end{multline}}

\newcommand{\senend}{\mbox{\; .}}
\newcommand{\sencon}{\mbox{\; ,}}
\begin{document}
\setlength{\tabcolsep}{1pt}
\title{Topological Insulator in the Presence of Spatially Correlated Disorder}
\author{Adrian Girschik}
\affiliation{Institute for Theoretical Physics, Vienna University of Technology, A-1040, Vienna, Austria, EU}
\author{Florian Libisch}
\affiliation{Department for Mechanical and Aerospace Engineering, Princeton University, Princeton, New Jersey 08544-5263, USA}
\author{Stefan Rotter}
\affiliation{Institute for Theoretical Physics, Vienna University of Technology, A-1040, Vienna, Austria, EU}
\begin{abstract}
  We investigate the effect of spatially correlated disorder on two-dimensional topological insulators and on the quantum spin Hall effect which the helical edge states in these systems give rise to. Our work expands the scope of previous investigations which found that uncorrelated disorder can induce a nontrivial phase called the topological Anderson insulator (TAI). In extension of these studies, we find that spatial correlations in the disorder can entirely suppress the emergence of the TAI phase. We show that this phenomenon is associated with a quantum percolation transition and quantify it by generalizing an existing effective medium theory to the case of correlated disorder potentials. The predictions of this theory are in good agreement with our numerics and may be crucial for future experiments. 
\end{abstract}
\maketitle
\section{Introduction}
\begin{figure}
  \begin{center}
    \includegraphics[angle=0, scale=1.0, width=0.5\textwidth]{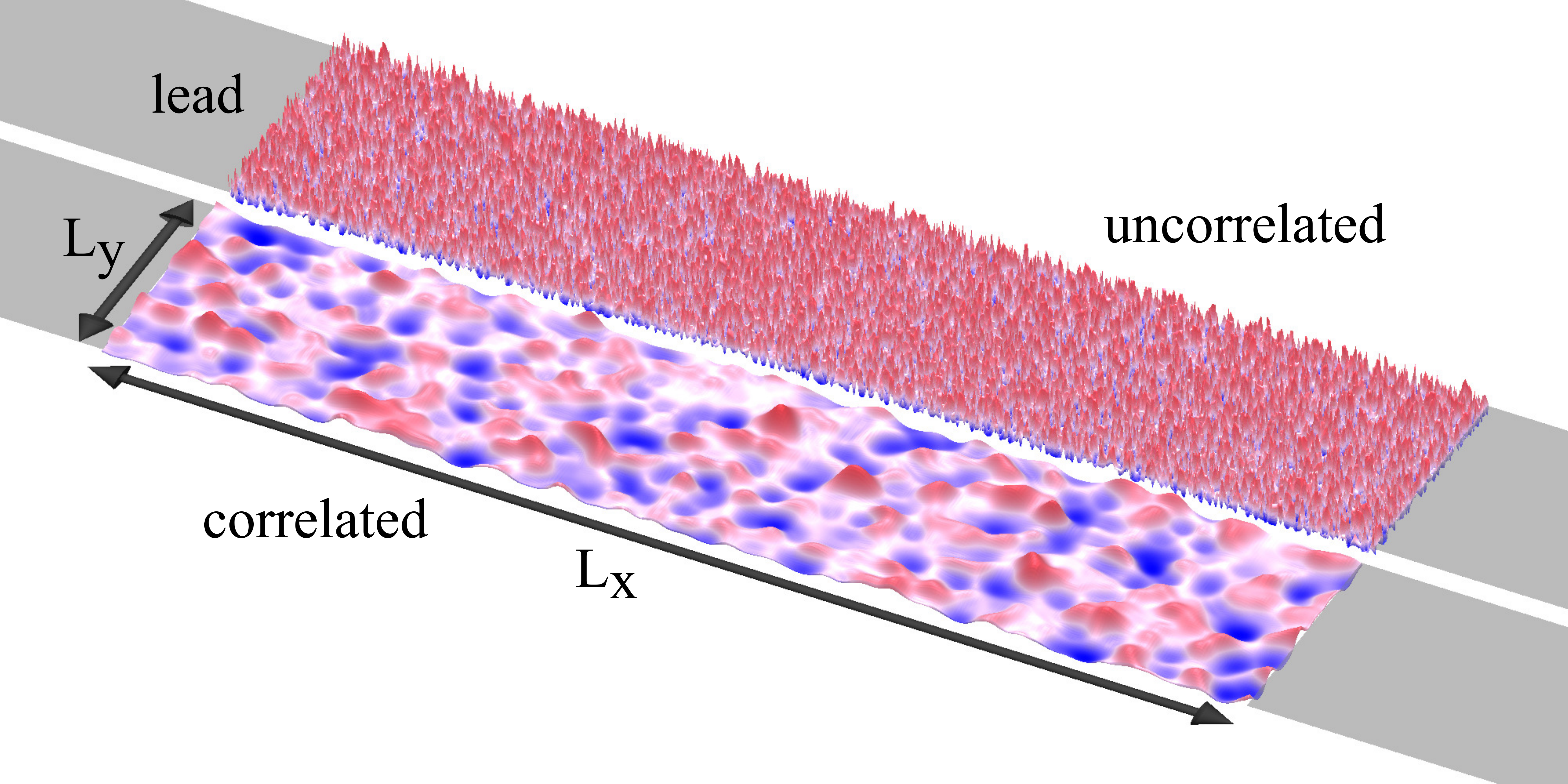} 
    \caption{(Color online). Illustration of the scattering setup for the considered topological insulator: A rectangular disordered middle part of height $L_y$ and width $L_x$ is attached to two semi-infinite leads on the left and right. Random potentials with and without correlations between neighboring grid points in the underlying lattice discretization are shown in the bottom and top panel, respectively.} 
    \label{fig:potentials}
  \end{center}
\end{figure}

A two-dimensional topological insulator \cite{HasKan2010} features edge states similar to those of the quantum Hall effect with the difference that electrons of different spin move in the same direction at opposite edges. Accordingly, this so-called ``quantum spin Hall'' (QSH) effect \cite{KanMel2005_2} can be understood as two noninteracting copies of a quantum Hall system, one for each spin. The stability of the edge states is guaranteed by time-reversal symmetry which forbids scattering into the counter-propagating edge state with opposite spin. \cite{MooBal2007,QiHugZha2008} These properties, which have recently attracted considerable attention, \cite{QiZha2010,Zha2008,KanMel2006,Bru2010,Fra2010,Moo2010} make topological insulators promising candidates for key components in future spin-tronic devices. \cite{MurNagZha2003,Bue2009} In 2006 HgTe/CdTe quantum wells were proposed as suitable systems for a first experimental realization of the QSH effect \cite{BerHugZha2006} which was, indeed, achieved shortly thereafter.\cite{KoeWieBru2007,RotBruBuh2009} Numerically it was found that the edge states in a topological insulator not only show great robustness against disorder but also that strong disorder itself can induce a phase featuring pure edge transport even if the system is an ordinary insulator in the clean limit.\cite{LiChuJai2009} This disorder-induced and topologically nontrivial phase was named topological Anderson insulator (TAI). In 2009 a theory was put forward \cite{GroWimAkh2009} that lead to a detailed understanding of the TAI, showing that disorder causes a negative correction to the topological mass which pushes the system into the TAI phase. This interesting phenomenon has meanwhile been investigated numerically in a variety of different systems \cite{XinZhaWan2011,YamNomImu2011,Pro2011,ZhaChuZha2012,SonLiuJia2012} including the case of a three-dimensional topological insulator. \cite{GuoRosRef2010} However, due to the challenges involved in controlling the disorder in a HgTe/CdTe quantum well, the TAI has not yet been realized experimentally. This problem might be overcome by employing ultra-cold atomic gases in optical lattices for the realization of a topological insulator.\cite{BerCoo2011} In such a highly tunable model system the disorder could be introduced by an optical laser speckle potential \cite{BilJosZuo2008,LyeFalMod2005} which has the advantage of being under external control. \\
An important point to emphasize in this context is that both the speckle pattern for cold atomic gases as well as the disorder which naturally occurs in a quantum well are characterized by a finite spatial correlation length $\xi$. Since this correlation has been disregarded in all previous numerical studies of the TAI which we are aware of \cite{LiChuJai2009,GroWimAkh2009,JiaWanSun2009,LiZanJia2011,SonLiuJia2012} the question was posed \cite{LiZanJia2011} how a finite correlation length $\xi$ would influence the predictions for the appearance and for the stability of the TAI. In view of the fact that spatial correlations in the disorder have already been shown to play an important role in the context of various other scattering scenarios \cite{MouLyr1998,CarBerIva2002,MeyNes2006,KawOnoOht2007,BarTwoBro2007,SchSchTom2008,PilGioPro2009,LugAspSan2009,LieCao2010,RotAmbLib2011,LiHwaRos2011,SedKesSar2011,IzrKroMak2012,CroSch2012,DieStoKuh2012,PetSan2012} one may expect such correlations to be a relevant factor also for topological insulators. We address this topic by studying explicitly how a static and spatially correlated disorder influences the transport characteristics of topological insulators (Fig \ref{fig:potentials}). As we will specify in detail below, our numerical results show marked deviations from conventional simulations with uncorrelated disorder.
\section{Method}
We proceed along the lines of previous studies, where an appropriate description of the two dimensional HgTe/CdTe quantum well was proposed in terms of an effective Hamiltonian.\cite{BerHugZha2006} This Hamiltonian which was derived based on the ${\mathbf k} \cdot {\mathbf p}$ perturbation theory and the six-band Kane-model takes the following form:
\begin{equation} \label{effective hamiltonian}
H_{\rm{eff}} (k_x,k_y)=
\left(
\begin{array}{cc}
h(\vec k) & 0 \\
0 & h^*(-\vec k)
\end{array} 
\right) \sencon
\end{equation}
with 
\begin{align}
h(\vec k)&=\mathds{1} \epsilon(\vec k) + d_i(\vec k) \sigma^i \label{h(k)} \\
\epsilon(\vec k)&=C-D\left(k_x^2+k_y^2\right)  \nonumber  \\
d_i&=\left( \begin{array}{c} Ak_x \\ -Ak_y \\ M(\vec k) \end{array}\right) \nonumber \\
M(\vec k) &= m-B(k_x^2 + k_y^2) \nonumber
\end{align}
and $\sigma^i$ labeling the Pauli-matrices. The basis of this effective Hamiltonian consists of the s-like E1 and the p-like heavy-hole H1 quantum well sub-bands for spin up ($+$) and down ($-$). The ordering is chosen to be $|E1+\rangle, |H1+\rangle, |E1-\rangle, |H1-\rangle$. Since the spin-up and spin-down parts in the Hamiltonian are decoupled as a consequence of time reversal symmetry \cite{ZhoLuChu2008} it is sufficient for our calculations to only use the spin-up block $h(\vec k)$. The solution for the spin-down block follows from a time-reversal operation. The material-dependent constants $A$, $B$ and $D$ in all our calculations are set to realistic values $A=364.5 \;\rm{meV}\,\rm{nm}$, $B=-686 \;\rm{meV} \, \rm{nm}^2$, $D=-512\; \rm{meV} \; \rm{nm}^2$ and $C=0\;\rm{meV}$ taken from Ref.~\onlinecite{KoeBuhMol2008}. The sign of the topological mass $m$ has a strong impact on the system's transport behavior: for positive $m$ the system behaves like an ordinary insulator with a band gap of $2\left|m\right|$, whereas if $m$ is set to a negative value the system turns into a topological insulator featuring perfectly transmitting edge states for the Fermi energy $E_F$ lying inside the bulk band gap $\left| E_F \right| < \left| m\right|$. To simulate such a system we use the experimentally determined value \cite{KoeBuhMol2008} of $m=-10\;\rm{meV}$.

The scattering geometry which we consider consists of a rectangular disordered region of width $L_x$ and height $L_y$ attached to two clean, semi-infinite leads (see illustration in Fig.~\ref{fig:potentials}).  We discretize the scattering region on a square lattice with grid constant $a$ using the tight-binding approximation in the continuum limit for the implementation of the effective Hamiltonian.  If not stated otherwise, the grid-constant $a$ is set to $5\; \rm{nm}$, in agreement with the value used in previous studies. \cite{LiChuJai2009,GroWimAkh2009,JiaWanSun2009,CheLiuLin2012,XuQiLiu2012} For simplicity, we consider the limit of vanishing temperature $T\to 0$ and infinitely small bias voltages $V\to 0$ applied between the two semi-infinite leads. According to the Landauer-B\"uttiker formalism the conductance $G$ in this limit is proportional to the transmission probability $T$ at the Fermi energy $E_F$, 
\begin{equation} \label{landauer-buttiker}
G=\frac{e^2}{h} T  = \frac{e^2}{h} \sum_{n,m}^N \left|t_{nm}\right|^2 \senend
\end{equation}
The indices $n$ and $m$ extend over all $N$ lead modes and $t_{nm}$ labels the transmission amplitude from mode $n$ in the incoming lead to mode $m$ in the outgoing lead. Since we consider both spins separately, every mode only contributes a single conductance quantum $e^2/h$. For the calculation of the transmission we employ the advanced modular recursive Green's function method \cite{RotWeiRoh2003,RotTanWir2000,LibRotBur2012} which incorporates the disorder by way of a static on-site energy value $V(\vec{x})$ imposed at every grid-point $\vec{x}=(x_i,y_j)$. In previous calculations the random on-site energies $V(\vec{x})$ were chosen to be uniformly distributed within a given energy interval $[-U/2,U/2]$, and each random sample from this distribution was drawn independently for each grid point. Since in this case the disorder value on each grid point has no correlation with the values on neighboring grid points we will refer to this type of disorder as ``uncorrelated''. To go beyond this limitation and to account for the spatial correlations which naturally occur in realistic situations we choose our disorder potential such as to obey the Gaussian correlation function
\begin{equation}
C(\vec{r})=\langle V(\vec{x}) \cdot V(\vec{x}+\vec{r}) \rangle \propto \exp\left({-\frac{r^2}{2 \xi^2}}\right) \label{correlation function}
\end{equation}
where the brackets $\langle ... \rangle$ stand for an average over all the grid points $\vec{x}$ and many disorder realizations. The standard deviation of this Gaussian defines the correlation length $\xi$, which measures the spatial range of the correlations. The value of the disorder strength $U$ is established by demanding 
\begin{align}
\langle V_{ij} \rangle = 0 \sencon \nonumber \\
\langle V_{ij}^2 \rangle = \frac{U^2}{12} \senend \label{moments} 
\end{align} 
These values are chosen such as to agree with those of the uncorrelated disorder potential distributed within the interval $[-U/2,U/2]$. See Fig.~\ref{fig:potentials} for an illustration of the disorder potentials with and without spatial correlations in the employed tight-binding grid.
\section{Results}
\begin{figure}
  \begin{center}
    \includegraphics[angle=0, scale=1.0, width=0.45\textwidth]{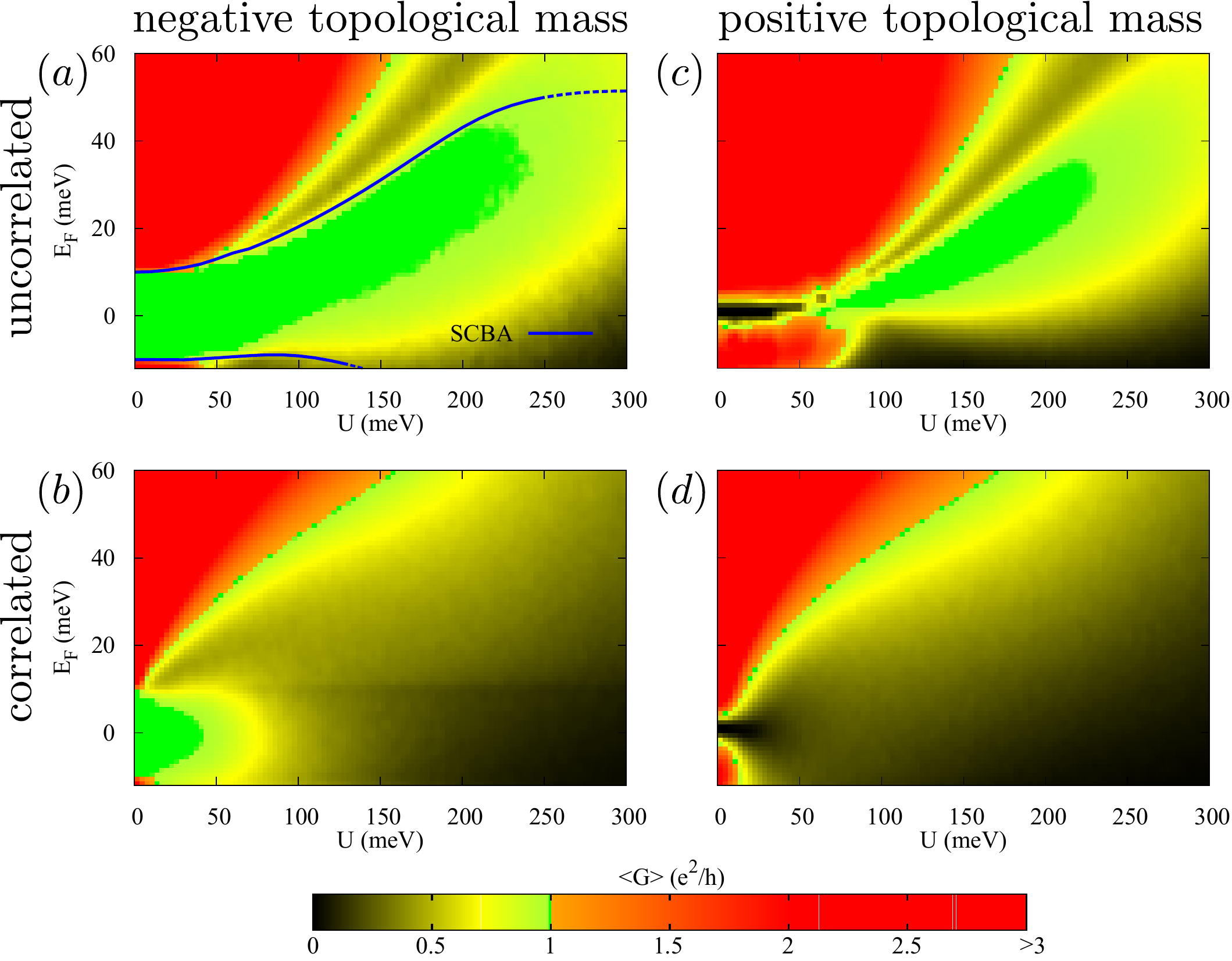}
    \caption{(Color online). Average conductance $\langle G \rangle$ as a function of disorder strength $U$ and Fermi energy $E_F$ for systems with negative $m=-10 \; \rm{meV}$ (left column) and with positive $m=1\;\rm{meV}$ (right column).  The system length $L_x=2000\;\rm{nm}$ and the height $L_y=500\;\rm{nm}$. The average is taken over $200$ in (a) and $1000$ disorder realizations in (b), (c) and (d). The green area shows where the average conductance $\langle G \rangle$ reaches a plateau at a single conductance quantum $e^2/h$ originating from edge transport through the disordered system. Top row: For uncorrelated disorder the TAI phase appears in this green area for strong disorder at energies where no edge states exist for $U=0$. The blue lines in (a) show the phase boundaries of the TAI predicted by the effective medium theory which is based on the self-consistent Born approximation (SCBA) shown in Eq. (\ref{self energy}). Bottom row: Spatial correlations in the disorder with correlation length $\xi=23.45\;\rm{nm}$ destroy the TAI conductance plateau and for $m<0$ also narrow the QSH plateau to an interval within the original bulk band gap. For positive topological mass (right column) the leads have been doped resulting in an energy offset of $\Delta=20\;\rm{meV}$ since otherwise no lead states would exist in the band gap.} 
    \label{fig:TAI_m-10}
  \end{center}
\end{figure}

We first consider the conductance through a disordered rectangular bar of width $L_x=2000\;\rm{nm}$ and height $L_y=500\;\rm{nm}$ for a negative and a positive value of the topological mass $m$ ($m=-10\;\rm{meV}$ and $m=+1\;\rm{meV}$), respectively. In the clean limit the system with $m<0$ [see $U=0$ in Fig.~\ref{fig:TAI_m-10}(a)] features quantized edge transport (green area) within the bulk band gap $\left|E_F\right|<\left|m\right|$, whereas conductance is entirely suppressed in the energy range $\left|E_F\right|<\left|m\right|$ for $m>0$ [see $U=0$ in Fig.~\ref{fig:TAI_m-10}(c)]. \cite{LiChuJai2009} Adding now an {\it uncorrelated} disorder to the clean systems with $m<0$ and $m>0$ gives rise to an unconventional conductance plateau [see $U>0$ in Figs.~\ref{fig:TAI_m-10}(a) and ~\ref{fig:TAI_m-10}(c) as well as Ref.~\onlinecite{LiChuJai2009}]. This so-called TAI phase of quantized transport emerges in the presence of strong uncorrelated disorder at energies at which no edge transport is present in the clean limit $U=0$. \cite{LiChuJai2009} In the case of $m<0$ (Fig.~\ref{fig:TAI_m-10}(a)) this TAI phase extends the original QSH phase beyond the disorder-free limits, given by $\left|E_F\right|=\left|m\right|$. Note that our results from Figs.~\ref{fig:TAI_m-10}(a) and ~\ref{fig:TAI_m-10}(c) agree very well with the literature,\cite{LiChuJai2009,GroWimAkh2009} thereby confirming the validity of our simulations. \\
In a next step we repeat this calculation for a {\it correlated} disorder potential. We choose the value of the correlation length $\xi=23.45\;\rm{nm}$ considerably larger than the grid constant $a=5 \;\rm{nm}$ but still much smaller than the height $L_y=500 \;\rm{nm}$ of the sample. Our results for such a finite correlation length [see Fig.~\ref{fig:TAI_m-10}(b) and \ref{fig:TAI_m-10}(d)] differ dramatically from the uncorrelated case [see Fig.~\ref{fig:TAI_m-10}(a) and \ref{fig:TAI_m-10}(c)]: Apparently the chosen spatial correlations in the disorder lead to a total breakdown of the TAI conductance plateau. In the case of $m<0$ we also observe an increased disorder-sensitivity as well as a narrowing of the conductance plateau corresponding to the original QSH phase in the clean limit. These results demonstrate that spatial correlations in the disorder add an important new component to the physics of topological insulators. Especially in view of the envisioned experiments that probe the physics of strongly disordered topological insulators, our results can apparently be expected to impose rather strict limits on the observability of the TAI phase. For such an experimental realization of the TAI we can certainly conclude that it is equally important to be able to control the correlation length $\xi$ as it is to control the strength $U$ of a disorder potential.

To check the influence of spatial correlations also for larger samples than the ones considered above we performed additional calculations. This is particularly important as finite-size effects in small samples due to a coupling of counter-propagating edge states can considerably distort the phase diagram of the TAI.\cite{ZhoLuChu2008,JiaWanSun2009,LiZanJia2011} To determine these phase boundaries in extended samples we performed a scaling analysis following previous work in this direction.\cite{YamNomImu2011,CheLiuLin2012} For this purpose a quadratic geometry of size $L=L_x=L_y$ is rolled up to a cylinder \cite{GroWimAkh2009,JiaWanSun2009,CheLiuLin2012} using periodic boundary conditions in the $y$ direction which eliminate the edge states in the sample. The disorder-averaged logarithmic conductance $\langle \ln G \rangle$ of the remaining bulk states is then calculated for three different sizes $L_1<L_2<L_3$ of the quadratic system as a function of the disorder strength $U$. This analysis allows us to estimate whether the bulk system in the limit of infinite size becomes conducting or insulating. For those values of $U$ where $\langle \ln G\rangle$ increases with increasing system size $L$, bulk states also conduct in an infinite system and thus suppress any kind of TAI phase due to the coupling of the edge states via bulk states. In contrast, in those regions where $\langle \ln G \rangle$ decreases with increasing system size $L$ the bulk is insulating in an infinitely large sample and clean edge transport can occur. The borders of these transitions between conducting and insulating bulk states can be estimated from the crossing points of $\langle \ln G\rangle$. \cite{CheLiuLin2012,XuQiLiu2012} We calculated these phase transition points for an uncorrelated and for correlated potentials with different correlation length $\xi$ in systems of three different sizes $L_1=500\;\rm{nm}$, $L_2=700\;\rm{nm}$ and $L_3=1050\;\rm{nm}$. Due to the high numerical effort involved, we restrict ourselves to a single energy of $E_F=16\;\mbox{meV}$ at which the TAI conductance plateau in the uncorrelated case is wide and well established [see Fig.~\ref{fig:TAI_m-10}(a)]. 

\begin{figure}
  \begin{center}
    \includegraphics[angle=0, scale=1.0, width=0.45\textwidth]{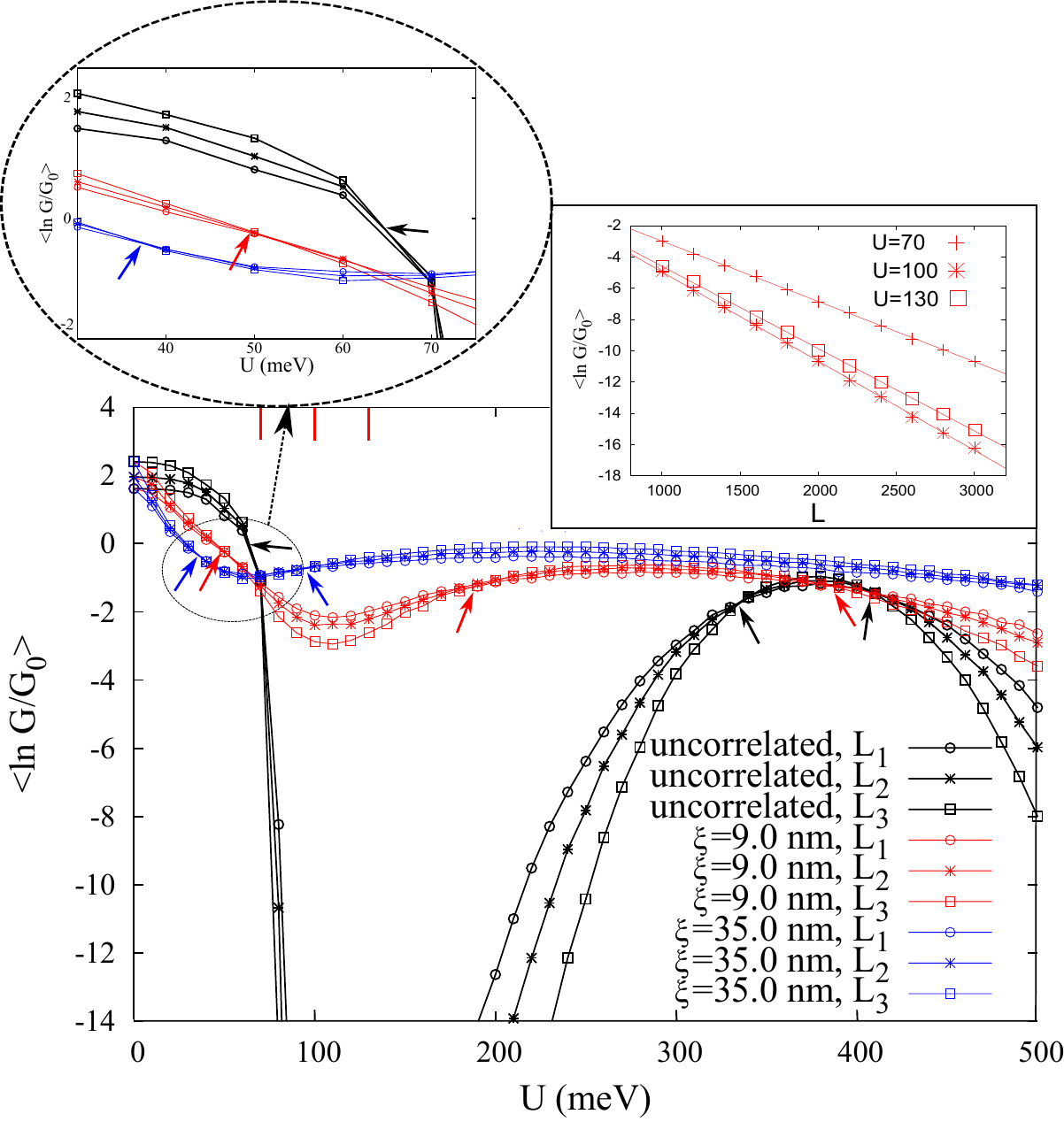} 
    \caption{(color online). Average logarithmic conductance $\langle \ln G\rangle$ through a rolled up quadratic topological insulator ($m=-10\;\mbox{meV}$) of size $L_i\times L_i$ as a function of disorder strength $U$. We consider three different system sizes $L_1=500 \; \mbox{nm}$, $L_2=700\;\mbox{nm}$ and $L_3=1050\;\mbox{nm}$ for uncorrelated disorder (black) and for correlated disorder with correlation length $\xi=9.0 \;\mbox{nm}$ (red) and $\xi=35.0\;\mbox{nm}$ (blue) at an energy of $E_F=16\;\mbox{meV}$ (disorder average taken over $2000$ configurations). The transitions into and out of the TAI phase occur at the crossing points of curves with equal color (their positions are marked by small arrows.) The TAI bulk band gap, present in the uncorrelated case in between the first and second crossing point, disappears for correlated potentials and the delocalization-localization region between second and third crossing point is broadened. The top right inset shows the behavior of $\langle \ln G\rangle$ as a function of the system length $L$ with fixed system width $W=500\,\mbox{nm}$ for $\xi=9\;\mbox{nm}$ and for three values of disorder strength $U=70$, $100$ and $130\;\mbox{meV}$ (see red vertical bars in the main panel). The localization length $L_{\mbox{loc}}$ is calculated from the slope of the best linear fit of $\langle \ln G\rangle$: $L_{\mbox{loc}}(U=70\,\mbox{meV})=518\,\mbox{nm}$, $L_{\mbox{loc}}(U=100\,\mbox{meV})=350\,\mbox{nm}$ and $L_{\mbox{loc}}(U=130\,\mbox{meV})=379\,\mbox{nm}$. The top left inset shows a closeup of some crossing points shown in the main panel.}
    \label{fig:scaling_analysis}
  \end{center}
\end{figure}

The results of our scaling analysis are shown in Fig.~\ref{fig:scaling_analysis}, where the uncorrelated case and two different correlation lengths $\xi$ are considered. For each of these three cases the dependence of $\langle \ln G\rangle$ on the disorder amplitude $U$ is shown. The phase transition points occurring at the crossing points of curves for different system sizes are marked by arrows. The top left inset of Fig.~\ref{fig:scaling_analysis} shows a closeup of the crossing points. As found previously, the crossing points move slightly to lower values of $U$ for increasing system size as a result of finite size effects. \cite{XuQiLiu2012} We can thus expect the real phase transition point to be to the left of our best estimate that we gain from the crossing points between the curves for $L_2$ and $L_3$ which are marked by the arrows in Fig.~\ref{fig:scaling_analysis}. Considering first the uncorrelated case studied already earlier (see black curves and arrows) we find that the lowest crossing point occurs at a value of the disorder strength $U\approx 65 \; \rm{meV}$ which fits well with the onset of the aforementioned TAI conductance plateau in Fig.~\ref{fig:TAI_m-10}(a) (see also Ref.~\onlinecite{CheLiuLin2012}). This onset coincides here with the opening of a bulk band gap which is reflected in the scaling plots of Fig.~\ref{fig:scaling_analysis} through a dramatic reduction of the conductance by more than ten orders of magnitude. The second and third crossing points, in turn, can be associated with the breakdown of the TAI phase observed in Fig.~\ref{fig:TAI_m-10}(a). The corresponding transition is, however, not induced by a band edge but rather by a mobility edge associated with those bulk states that fill the band gap when increasing the disorder strength beyond the first crossing point. These bulk states undergo a delocalization-localization transition at the second and third crossing point which destroys the conductance plateau as soon as the delocalized bulk states start coupling the edge states at opposite edges in the sample. Note that for this to happen it is already sufficient for individual rather than for all bulk states to delocalize such that finite size effects do play a role at this strong-disorder boundary of the TAI.\cite{CheLiuLin2012,XuQiLiu2012} 

\begin{figure} 
  \begin{center}
    \includegraphics[angle=0, scale=1.0, width=0.45\textwidth]{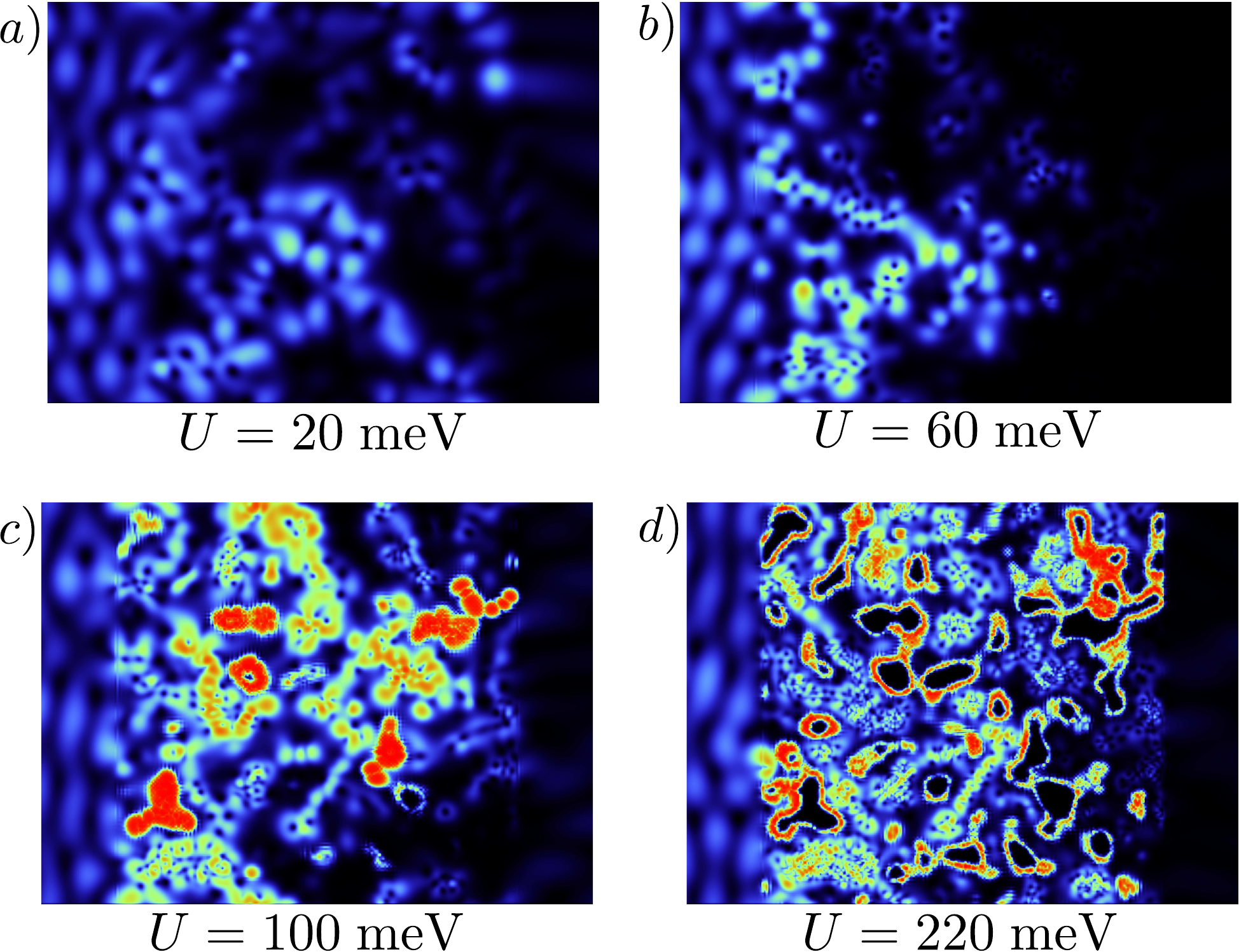} 
    \caption{Scattering wave functions $\left|\psi\right|^2$ in a cylindrical system of length $L=W=1050\;\mbox{nm}$, Fermi energy $E_F=16\;\mbox{meV}$ in a random potential of correlation length $\xi=35\;\mbox{nm}$. The flux is incoming from the left and periodic boundary conditions are implemented on the top and bottom of the images. The disorder strength $U$ for each of the pictures shown is indicated right below each panel. Note how the wave function turns into a percolating state when the localization-delocalization transition is approached for increasing disorder strength $U$ (compare with blue curves in Fig.~\ref{fig:scaling_analysis} and with experimental data from quantum Hall measurements as in Fig.~\ref{fig:TAI_m-10} of Ref.~\onlinecite{HasSohWie2008}). }
    \label{fig:percolation}
  \end{center}
\end{figure}

\begin{figure}
  \begin{center}
    \includegraphics[angle=0, scale=1.0, width=0.45\textwidth]{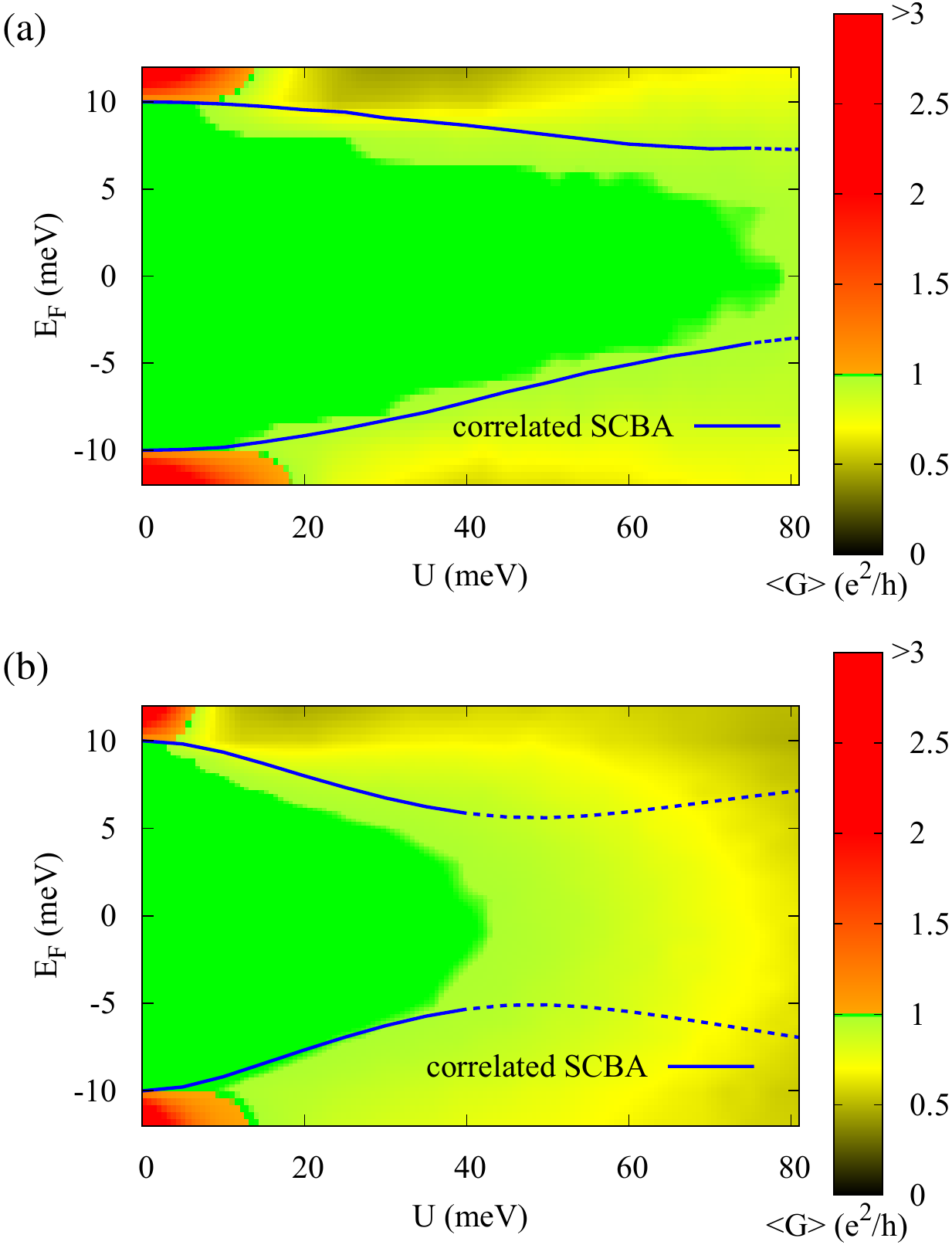}
    \caption{(Color online). The average conductance $\langle G \rangle$ as a function of Fermi energy $E_F$ and disorder strength $U$ is shown for the system considered in Fig.~\ref{fig:TAI_m-10}(b), here with correlation length (a) $\xi=9 \; \rm{nm}$ and (b) $\xi=23.45 \; \rm{nm}$. The system is $L_x=2000 \; \rm{nm}$ long and $L_y=500\;\rm{nm}$ high while the grid spacing $a=5\;\rm{nm}$. The blue curves delineate the borders of the quantized conductance plateau as estimated by the effective medium theory for correlated potentials, Eq. (\ref{self energy corr}). Note the very good agreement which we find with our numerical results.} 
    \label{fig:TAI_m-10_length}
  \end{center}
\end{figure}

When extending the above scaling analysis now to the case of correlated disorder with successively increasing correlation length $\xi$ we find a behavior different from the uncorrelated case: Already for the case of very short-range correlations with $\xi=9.0 \; \mbox{nm}$ (red curves in Fig.~\ref{fig:scaling_analysis}) significant differences appear. We still find three crossing points as before, but the conductance no longer displays the very strong suppression associated with a bulk band gap. Instead, we find that the delocalization-localization region of bulk states, which was previously associated with the strong-disorder boundary of the TAI, widens for increasing correlation length $\xi$. Correspondingly, in the disorder interval between the first and the second crossing points (which are now also much closer together) the conductance is much less suppressed than in the uncorrelated case. This indicates that in the correlated case the bulk band gap disappeared and was filled with localized bulk states. \\
To prove this statement we investigate more closely the behavior of $\langle \ln G\rangle$ in the disordered cylinder with surface area $W\times L$. Keeping the circumference of the cylinder and the correlation length of the potential fixed at $W=500\; \mbox{nm},\,\xi=9.0\;\mbox{nm}$, we vary the system length $L$ and consider three different values of $U$ within the region between the first and second crossing point where the bulk system is insulating in an infinitely large system. The results are shown in the top right inset of Fig.~\ref{fig:scaling_analysis}. The bulk states are indeed localized as the averaged logarithmic conductance $\langle \ln G\rangle$ drops linearly with increasing length $L$. From the slope $k_l$ of the fitted lines we determine the localization length $L_{\mbox{loc}}=-2/k_l$ of the bulk states which ranges from $518\;\mbox{nm}$ right after the first phase transition point ($U = 70\;\mbox{meV}$) down to $349\; \mbox{nm}$ in the middle of the ''insulating'' region ($U=100\;\mbox{meV}$). With the localization length $\xi$ thus falling below the linear dimension $W=L$  of the quadratic disorder region considered in Fig.~\ref{fig:scaling_analysis}, we can understand that the reduced bulk conductance is here produced by the localization of bulk states, rather than by a band edge as in the uncorrelated case.
We emphasize, however, that both a band gap as well as localized bulk states can give rise to a TAI, as was explicitly pointed out in a recent study:\cite{XuQiLiu2012} In what was termed a TAI-I phase the coupling between edge states is prevented by a bulk band gap which eliminates all bulk states that would mediate such a coupling. In a system with negative topological mass $m<0$ the TAI-I conductance plateau is joined with the original QSH plateau existing within the original bulk band gap $E_F<\left|m\right|$. A second TAI-II phase was characterized by a coexistence of localized bulk-states and extended edge-states. As long as the localization length of these bulk states remains smaller than the width of the sample, the coupling of edge states remains suppressed and the TAI persists. Following these arguments, the transition into the TAI phase can either occur at a band edge (for TAI-I) or at a mobility edge (for TAI-II). From this we conclude that the original band edge at the weak-disorder boundary of the TAI-I phase in the uncorrelated case gets replaced by a mobility edge as the new weak-disorder boundary of a TAI-II phase in the correlated case. The corresponding suppression of the TAI-I phase already at a rather small correlation length of $\xi=9.0 \; \rm{nm}$ suggests that the TAI-II phase is more robust to spatial disorder correlation than the TAI-I phase. The TAI-II phase is, in turn, more sensitive to finite size effects due to {\it individual} localized bulk states which can couple counter-propagating edge states to each other. Correspondingly we can understand the absence of a TAI conductance plateau in Fig.~\ref{fig:TAI_m-10}(b) and Fig.~\ref{fig:TAI_m-10_length} in between the first and second crossing points of our scaling analysis in Fig.~\ref{fig:scaling_analysis} as a finite-size effect which may disappear for much larger samples than studied here. Further explicit calculations will be necessary to better understand the infinite-size limit for TIs with long-range correlations in the disorder. Our own results for the case of $\xi=35 \; \rm{nm}$ (see the blue curves in Fig.~\ref{fig:scaling_analysis}) show that the widening of the delocalization-localization transition continues for increasing correlation length $\xi$. However, since the transition region for $\xi=35 \; \rm{nm}$ is here already very wide, detailed statements on the phase boundaries in the infinite size limit are difficult to deduce from our finite-size calculations. \\
We may, however, get important insights into the nature of the localization-delocalization transition for correlated disorder potentials by explicitly studying the scattering wave functions close to this transition. In Fig.~\ref{fig:percolation} we plot several such wave functions for increasing disorder strength $U$ in our cylindrical system of size $L = W = 1050\;\mbox{nm}$ and correlation length $\xi = 35\;\mbox{nm}$ (compare with blue curves in Fig.~\ref{fig:scaling_analysis}). These plots indicate that the observed localization-delocalization transition is, in fact, a percolation transition similar to the one in the quantum Hall effect. \cite{Huc1995} At the percolation threshold which is realized at critical values of the system parameters (like the disorder strength $U$) localized bulk states turn into extended states which circumnavigate the hills and valleys of the disorder potential rather than being trapped by them. The wave functions shown in Fig.~\ref{fig:percolation} indicate exactly such a behavior by displaying bulk states that propagate along the slopes of pronounced potential variations [see Fig.~\ref{fig:percolation}(d)] as observed in quantum Hall measurements (see Fig.~2 in Ref.~\onlinecite{HasSohWie2008}). This percolation explains the suppression of uni-directional edge transport quite intuitively since close to the percolation threshold the bulk states which are otherwise localized may percolate from one edge to the opposite one and thereby couple the counter-propagating edge states. Note that our observation of the percolation transition fits well with earlier work \cite{GroWimAkh2009} that found the critical exponent for this transition in the uncorrelated case to be consistent with the exponent from the quantum Hall universality class. The most closely related work to this paper which we could identify is by Shen \it{et al.} \cite{ChuLuShe2012} who recently demonstrated that bound states in a quantum spin Hall anti-dot lattice feature a percolation transition {\it in the bulk band gap}. Since in the present system we observe the percolation transition at $E_F=16$ meV, which is well outside the bulk band gap at $\left|E_F\right|<10$ meV, a different mechanism seems to be at work here which we intend to discuss in a separate paper. \cite{Gir20XX} \\
In the following we will present additional evidence to corroborate our arguments from above with respect to the suppression of the TAI-I phase due to correlations in the disorder. If these arguments are correct, the QSH phase (characterized by a negative topological mass and a chemical potential in the band gap, $|\mu|<|m|$) should border,  in the correlated disorder case, directly on the surviving TAI-II phase (characterized by a negative topological mass and a chemical potential outside of the band gap, $|\mu|>|m|$). In the corresponding plot in Fig.~\ref{fig:TAI_m-10}(b) we see that the QSH conductance plateau (existing in the energy-range $-10\;\rm{meV}<\mu<10\;\rm{meV}$ in the disorder-free sample) extends only to much smaller values of disorder strength $U$ than in the uncorrelated case (compare with Fig.~\ref{fig:TAI_m-10}(a)). Since for the above arguments the borders of this reduced QSH conductance plateau with the neighboring TAI-II phase are characterized by a band edge crossing, these borders should be describable in terms of a similar effective medium theory as has been developed for the uncorrelated case.\cite{GroWimAkh2009} This theory maps the disordered system onto a disorder-free sample with a renormalized topological mass $\bar{m}=m+\delta m$ and chemical potential $\bar{\mu}=\mu+\delta \mu$. This renormalization was carried out in terms of the self-consistent Born approximation (SCBA) using an integral equation for the self-energy $\Sigma$:
\begin{align}
\Sigma= \frac{U^2}{12}&\left(\frac{a}{2\pi}\right)^2 \lim_{\kappa\to 0}\int_{-\frac{\pi}{a}}^{\frac{\pi}{a}}dk_xdk_y \nonumber \\
&\times \left(E_F+i\kappa-H_0(\vec k)-\Sigma\right)^{-1}  \label{self energy}\senend 
\end{align}
Whenever the renormalized chemical potential reaches the edge of the band gap ($|\bar{\mu}|=|\bar{m}|$) the border of the QSH (or TAI-I) phase has been reached. Since this indicator, as calculated through the above SCBA, is independent of the system size, the effective medium theory offers an insightful and practical tool to determine the boundaries of the QSH or possible TAI-I phases in the infinite-size limit. 

In order to generalize the effective medium theory from above to the case of spatially correlated potentials we resort to recent theoretical work in which an extension of the coherent potential approximation to correlated disorder was proposed. \cite{ZimSch2009} Following this line of work, one can conveniently include the disorder correlations through an additional term in Eq.~(\ref{self energy}) which is given by the normalized Fourier transform of the disorder correlation function, \footnote{Note that in our case the self-energy $\Sigma$ is independent of $\vec{k}$, such that correlations between different $\vec{k}$-vectors as originally described in Eq.~(27) of Ref.~\onlinecite{ZimSch2009} reduce here to the simplified expression in Eq.~(\ref{self energy corr})} 
 $\tilde{C}(\vec{k})$,
\begin{align}
\Sigma= \frac{U^2}{12}& \lim_{\kappa\to 0}\int_{-\frac{\pi}{a}}^{\frac{\pi}{a}}dk_xdk_y \nonumber \\
&\tilde{C}(\vec{k})\times \left(E_F+i\kappa-H_0(\vec k)-\Sigma\right)^{-1}  \label{self energy corr}\senend 
\end{align}
Since in the present case of Gaussian disorder correlations [see Eq.~(\ref{correlation function})] the expression for $\tilde{C}(\vec{k})$ is a Gaussian itself (centered around $k=0$), the effect of the disorder correlations is to smoothly cut off the above integral. The more long-range the correlations are (in real space), the sharper this cut-off is (in Fourier space). 

To extract the corrections $\delta m$ and $\delta \mu$ from the self-energy in Eq.~(\ref{self energy corr}), we proceed along the lines of the   uncorrelated case \cite{GroWimAkh2009} and decompose the self-energy into the Pauli matrices $\sigma_i$
\begin{equation} \label{decomposition self energy}
\Sigma = \Sigma_0 \sigma_0+\Sigma_x \sigma_x+\Sigma_y \sigma_y+\Sigma_z \sigma_z
\end{equation}
with the help of
\begin{align} 
\delta m=\mbox{Re} \;\Sigma_z \sencon \nonumber  \\
\delta \mu=-\mbox{Re}\; \Sigma_0 \senend \label{delta e and m}
\end{align} 
With the values extracted for the renormalized topological mass $\bar{m}=m+\delta m$ and chemical potential $\bar{\mu}=E_F+\delta \mu$, we can now estimate the boundaries of the QSH or TAI-I phase in the case of correlated disorder by determining the values of $E_F$ and $U$ for which the renormalized chemical potential $\bar{\mu}$ drops into the effective band gap at $\bar{\mu}=\pm \bar{m}$. We start by first testing our approach for the {\it uncorrelated} case, for which the weak disorder boundary of the TAI phase was estimated before. In this case the discrete Fourier transform of the correlation function $\tilde{C}(\vec{k})$ is constant. By normalizing this function in $k$-space to the volume $V=(2\pi)^2/a^2$ of the Brillouin zone, we exactly reobtain the expression for the self energy in the uncorrelated case, Eq.~(\ref{self energy}). If we determine with this approach the phase boundaries of the QSH and TAI-I phases in the {\it uncorrelated} case we obtain the blue curves in Fig.~\ref{fig:TAI_m-10}(a) which fit nicely to the conductance plateau of the QSH and TAI-I phases and to previous calculations. \cite{LiChuJai2009,GroWimAkh2009,CheLiuLin2012} Extending our calculations to the case of {\it correlated} disorder, the borders which we calculate through Eqs.~(\ref{self energy corr}) and (\ref{delta e and m}) (see blue curves in Fig.~\ref{fig:TAI_m-10_length}) describe the boundaries of the QSH conductance plateau very well (without any adjustable parameters). The good agreement which we find for different correlation lengths $\xi$ [see Fig.~\ref{fig:TAI_m-10_length}(a) and \ref{fig:TAI_m-10_length}(b)] corroborates the validity of our approach. Note that, in contrast to the uncorrelated case, no TAI-I conductance plateau is observed for $|\mu|>|m|$ (i.e., outside of the energy region where the QSH phase is present in the clean sample). An increasing disorder strength rather leads to a narrowing of the bulk band gap within which pure QSH edge transport can occur. This reduced band gap corresponds to positive corrections $\delta m$ and $\delta \mu$ in the case of a correlated potential whereas in the uncorrelated case these corrections were shown to be negative.\cite{GroWimAkh2009} It is exactly these positive corrections which lead to the breakdown of the TAI-I phase that occurs in the uncorrelated case for $|\mu|>|m|$.
\section{Conclusion}
 In this work we investigate the effect of spatially correlated disorder on a two-dimensional topological insulator. We thereby extend previous studies in which only uncorrelated disorder potentials were considered. \cite{LiChuJai2009,GroWimAkh2009,CheLiuLin2012,XuQiLiu2012} Our calculations show that a finite correlation of the disorder potential enhances finite-size effects and may entirely suppress a regime of quantized conductance known as the topological Anderson insulator phase. We link this phenomenon with a quantum percolation transition that we find to occur in the limit of correlated strong disorder (a detailed study on this will be published separately). \cite{Gir20XX} To describe the observed boundaries of quantized conductance theoretically, we perform a scaling analysis and adapt an existing effective medium theory to the case of spatially correlated potentials which yields quantitative agreement with our numerics. Our results suggest that for observing the topological Anderson insulator phase experimentally, it will be necessary to work with comparatively large samples (to suppress finite size effects) and with very short ranged disorder potentials as any long-range correlations may strongly suppress this topologically nontrivial phase. We speculate that spatial correlations might also be an important impediment to eliminate the bulk conductance in {\it three-dimensional} topological insulators.\cite{PenLaiKon2010} This would certainly constitute an interesting topic for further investigations.
\section{Acknowledgements}
We thank C. W. J. Beenakker and M. Wimmer for very helpful discussions. Our calculations were performed on the Vienna Scientific Cluster (VSC). 
\bibliography{TIcorr_submit}
\end{document}